\documentclass[universe,article,accept,pdftex,moreauthors]{Definitions/mdpi} 

\usepackage{bm}
\firstpage{1} 
\pubvolume{1}
\issuenum{1}
\articlenumber{0}
\pubyear{2025}
\copyrightyear{2025}
\externaleditor{Firstname Lastname}
\datereceived{ } 
\daterevised{ } 
\dateaccepted{ } 
\datepublished{ } 
\hreflink{https://doi.org/} 

	\Title{Neutron Star Inner Crust at Finite Temperatures: A Comparison Between Compressible Liquid Drop and Extended Thomas--Fermi Approaches}

	\TitleCitation{Neutron Star Inner Crust at Finite Temperatures: A Comparison Between Compressible Liquid Drop and Extended Thomas--Fermi Approaches}
	
	\Author{Guilherme Grams 
		$^{1,2,3,}$*\orcidA,
		Nikolai N. Shchechilin $^{1,2}$\orcidB,
		Th\'eau Diverr\`es 
		$^{4}$\orcidC, 
		Anthea F. Fantina $^{4,1,2}$\orcidD, \linebreak
		Nicolas Chamel $^{1,2}$\orcidE ~and Francesca Gulminelli $^{5}$\orcidF
	}
	
	\AuthorNames{Guilherme Grams, Nikolai N. Shchechilin, Th\'eau Diverr\`es, Anthea F. Fantina, Nicolas Chamel and Francesca Gulminelli}
	
	\isAPAStyle{%
		\AuthorCitation{Lastname, F., Lastname, F., \& Lastname, F.}
	}{%
		\isChicagoStyle{%
			\AuthorCitation{Guilherme Grams, Nikolai N. Shchechilin, Th\'eau Diverr\`es, Anthea F. Fantina, Nicolas Chamel and Francesca Gulminelli}
		}{
			\AuthorCitation{Grams, G.; Shchechilin, N.N.; Diverr\`es, T.; Fantina, A.F.; Chamel, N.; Gulminelli, F.}
		}
	}
	\address{$^{1}$ \quad Institut d'Astronomie et d'Astrophysique, CP-226, Université Libre de Bruxelles, 1050 Brussels, Belgium; nikolai.shchechilin@ulb.be (N.N.S.);  nicolas.chamel@ulb.be (N.C.) \\
		$^{2}$ \quad BLU-ULB, Brussels Laboratory of the Universe, 50 avenue F.D. Roosevelt, B-1050 Brussels, Belgium
		\\
		$^{3}$ \quad Institut für Physik und Astronomie, Universität Potsdam, Haus 28, Karl-Liebknecht-Str. 24/25, 14476~Potsdam, Germany\\
		$^{4}$ \quad Grand Acc\'el\'erateur National d'Ions Lourds (GANIL), CEA/DRF-CNRS/IN2P3, Boulevard Henri Becquerel, 14076 Caen, France; theau.diverres@ganil.fr (T.D.); anthea.fantina@ganil.fr (A.F.F.) \\
		$^{5}$ \quad Universit\'e de Caen Normandie, ENSICAEN,
		CNRS/IN2P3, LPC Caen UMR6534, 14000 Caen, France; gulminelli@lpccaen.in2p3.fr (F.G.)}
	
	\corres{Correspondence: guilherme.grams@uni-potsdam.de (G.G.) 
	}
	
	\abstract{ 
		We investigate the effects of temperature on the properties of the inner crust of a non-accreting neutron star.
		To this aim, we employ two different treatments: the compressible liquid drop model (CLDM) and the temperature-dependent extended Thomas--Fermi (TETF) method. 
		Our systematic comparison shows an agreement between the two methods on their predictions for the crust thermodynamic properties. We find that the CLDM description can also reproduce reasonably well the TETF composition especially if the surface energy is optimized on the ETF calculation. However, the neglect of neutron skin in CLDM leads to an overestimation of the proton radii.  
	}
	
	
\begin{document}

		\section{Introduction}
		\label{sec:intro}
		
		Neutron stars (NS) offer a unique laboratory to study matter under extreme conditions. 
		The rich structure of their inner crust consists of
		neutron--proton clusters surrounded by a sea of unbound nucleons and electrons.
		An accurate description of nuclear clusters in the NS crust is important, for example, for the determination of elastic properties 
		(governing the formation of `mountains', which are small asymmetrical deformations in the NS solid crust and can be a source of continuous gravitational-wave emission), transport coefficients, and neutrino propagation \cite{Chamel2008,Gittins2020,Foucart2023}.
		Since NSs are born hot from supernova core collapse, at least for the initial stages of their evolution, the properties of the crust and of the dense-matter equation of state (EoS) need to be investigated at finite temperature. 
		This opens the possibility to study proto-NSs and their cooling (see, e.g., Refs.~\cite{prakash1997, yakovlev2004, pons1999} for a review, and Refs.~\cite{dinh2023aa, Dehman2024, Scurto2024} for recent works). 
		Moreover, finite-temperature EoSs are essential for understanding transient events such as core-collapse supernova and NS mergers (see, e.g., Refs~\cite{Lattimer91,Oertel17,Schneider2017,Tsiopelas24}).
		
		Modeling the NS inner crust at finite temperature is a challenging task (see, e.g., Refs.~\cite{Oertel17,Burgio18,Burgio2021} for a review), especially within self-consistent mean-field-based methods such as Hartree--Fock, Hartree--Fock with Bardeen--Cooper--Schrieffer pairing, or Hartree--Fock--Bogoliubov (HFB) approaches. 
		For this reason, because of their simplicity and fast computation, most studies have been carried out within compressible liquid drop models (CLDM), with parameters fitted to experimental nuclear binding energies (see, e.g., Refs.~\cite{Lattimer91, Carreau2020a, dinh2023aa,Schneider2017}). 
		Alternatively, temperature-dependent (extended) Thomas--Fermi (T(E)TF) approaches, which provide a more microscopic (although semi-classical) description of nuclear clusters, including smooth density profiles, have been employed~\cite{Brack85,Bartel1985}.
		Some groups have used TTF \cite{Shen11}, TETF \cite{Onsi97}, or TETF with proton shell corrections added with the Strutinsky-integral (TETFSI) method \cite{Onsi08}.
		Nonetheless, to our knowledge, a systematic comparison of the temperature effects on NS crust properties obtained within TETF and CLDM approaches has not been performed so far.
		This is of particular interest as currently available finite-temperature EoSs for proto-NS, NS mergers, and core-collapse supernova simulations are based either on CLDM (see, e.g., Refs.~\cite{Lattimer91, hempel2010, Schneider2017, radgul2019, steiner2013, furusawa2013}) or TTF approaches~\cite{Shen11, Togashi2017}, thus neglecting higher order density gradient terms present in the TETF energy functional.
		Therefore, the present work also represents a first step towards the application of the TETF approach to the calculation of the EoS for hot dense matter. 
		The similarities and differences between the CLDM and the ETF approaches to model the NS crust at zero temperature were recently compared in Ref.~\cite{Grams_2022}.
		The outcome showed that thermodynamic quantities such as pressure, energy, and chemical potential are in good agreement among the models, while the details of the finite-size modeling mainly impact the composition of the nuclear clusters.
		Comparisons between CLDM and TETF calculations at finite temperature were later performed by the authors of Ref.~\cite{Carreau2020a}.
		They demonstrated, using the BSk14 functional~\cite{Onsi08}, that as temperature approaches the crustal melting point, the composition (and particularly the number of protons) in the TETF method is better reproduced by the CLDM if the surface energy is optimized on the ETF calculations corresponding to the same interaction. 
		
		In this work, we perform a more detailed and systematic comparison of the CLDM and the TETF approach, extending the studies of Ref.~\cite{Carreau2020a}, which focused on crystallization temperatures within the CLDM, and of Ref.~\cite{Grams_2022}, which compared CLDM and ETF at zero temperature. Here, we examine the finite-temperature properties of the NS inner crust. Namely, we analyze the temperature dependence of the thermodynamic quantities, the evolution of the proton and neutron numbers, as well as the nucleon density profiles in the Wigner--Seitz (WS) cell. 
		Moreover, we provide for the first time the EoS and constitution of the NS inner crust within the second-order TETF method. All calculations are based on the accurately calibrated BSk24 functional~\cite{Goriely13}, previously used to construct a unified EoS at zero temperature~\cite{pearson2018} and in finite-temperature CLDM studies~\cite{Carreau2020a, Carreau2020b, dinh2023aa, dinh2023aamcp}. The use of the same functional in both TETF and CLDM computations allows us to consistently address the discrepancies in the predictions of the inner-crust properties due to the use of different finite-size treatments, disentangling them from the uncertainties due to the choice of the~functional.
		
		The paper is organized as follows: we describe the formalism of both the TETF and the CLDM in Section~\ref{sec:method}, while we present our results in Section~\ref{sec:results}.
		To assess the robustness of CLDM predictions, we carry out three distinct surface-parameter fitting protocols, as also discussed in Appendixes~\ref{app:etf-fit} and \ref{sec:surf-en}.
		We finally draw our conclusions in Section~\ref{sec:concl}.

		
		\section{Treatment of the Inner Crust at Finite Temperature}
		
		\label{sec:method}
		
		In this section, we discuss the treatment of the inner crust at finite temperature in the TETF (see Section~\ref{sec:TETF}) and the CLDM (see Section~\ref{sec:cldm}) approaches.
		In both cases, nuclei are described within the one-component plasma (OCP) approximation, that is, at each density in the crust, the nuclear distribution that is expected to be present at finite temperature is represented by only one cluster, the one that is more favorable from the thermodynamic point of view.
		This assumption can be justified below the crystallization temperature, which for the inner crust is found to be $\sim$$0.1$--$1$~MeV \cite{haensel2007, Carreau2020a}, since the cluster distribution is relatively narrow, particularly in the shallowest layers. 
		In the temperature range of interest here, $k_{\rm B} T \lesssim 2$~MeV ($k_{\rm B}$ being the Boltzmann constant), the question of the importance of considering a full distribution of clusters may thus arise \cite{dinh2023aamcp}. 
		However, thermodynamic quantities are hardly affected \cite{burrows1984}, and it is, therefore, reasonable to ignore this distribution as a first step.
		In our OCP treatment, we also neglect the translational degrees of freedom associated with the center-of-mass motion of the clusters.
		The importance of the contribution of the translational free energy in the description of the finite-temperature crust was highlighted in Ref.~\cite{dinh2023aa}.
		However, for comparison and consistency with TETF calculations, we do not include this term in the CLDM treatment here.
		
		In addition, as shown in Ref.~\cite{dinh2023aa}, for the considered temperature regimes, $k_{\rm B} T \lesssim 2$~MeV, the proton--gas density remains very small, $10^{-3}$~fm$^{-3}$ at most at the bottom of the crust, and its effects on the crust EoS and composition are negligible.
		For these reasons, we ignore the presence of the free protons in the present work, as in Ref.~\cite{Carreau2020b}.
		
		We also assume clusters to be spherical. 
		Indeed, the so-called nuclear `pasta' phases are only supposed to be present at the bottom of the inner crust and are not expected to have a significant effect on the EoS. 
		In addition, while the presence of non-spherical clusters is usually predicted in CLDM and ETF approaches (see, e.g., Refs.~\cite{balliet2021, dinh2021aa,pearson2020,Parmar2022} for some recent works and Ref.~\cite{Lim2017,shchechilin2022} for a comparison between (E)TF and CLDM predictions), the inclusion of microscopic shell corrections seem to reduce the density range at which pasta phases appear \cite{pearson2022, Shchechilin23}.
		
		Moreover, all the calculations performed in this work are carried out in beta equilibrium, as achieved in late (proto-)NS cooling stages. 
		Indeed, below saturation density and at about $k_{\rm B} T$$\sim$$1$~MeV, beta equilibrium is likely to be attained (see, e.g., Refs.~\cite{pons1999, pascal2022, alford2021}).
		At higher temperatures, the situation is a priori less clear but estimates show that beta equilibrium is attained almost everywhere in the proto-NS during its evolution, apart from a thin region near the stellar surface at early times \cite{camelio2017}. 
		
		Finally, we rely on the widely used WS approximation~\cite{Wigner33,Wigner34,Baym71,Salpeter1961,Negele73,Bonche1981}, according to which the crust is decomposed into independent spherically symmetric and electrically charge-neutral cells around each cluster, whose volume $V_{\rm WS}$ contains $A_{\rm tot}$ nucleons and is defined such that the mean cluster number density is $1/V_{\rm WS}$. 
		This approximation introduces spurious shell effects in mean-field calculations~\cite{Chamel07,Newton2009}. 
		Such a limitation, however, does not arise for CLDM and ETF approaches.

		\subsection{The Temperature-Dependent ETF}
		\label{sec:TETF}
		
		We follow the TETF formalism for the clusterized matter in the hot NS inner crust, as was described in Ref.~\cite{Onsi97}. 
		The total free-energy density of the WS cell is expressed as
		\begin{equation}
			\mathcal{F}_{\rm TETF} = \mathcal{F}_e  + \mathcal{F}_{\rm Coul}
			+\mathcal{F}_{\rm nuc} \ , 
			\label{eq:ETF_total_free_energy_density}
		\end{equation}
		where $\mathcal{F}_e$ corresponds
		\endnote{We use the $F$ and $\mathcal{F}$ to denote the free energy per cell and the free energy per unit volume, respectively.} to the free-energy density of homogeneously distributed electron gas of number density $n_e$ and $\mathcal{F}_{\rm Coul}$ accounts for the Coulomb interactions. Considering the cell with proton density profile $n_p(r)$ and neglecting exchange contributions we write 
		\begin{equation}
			\mathcal{F}_{\rm Coul} = \frac{8\pi^2e^2}{V_{\rm WS}} \int_0^{r_{\rm WS}} \left(\frac{w(r)}{r}\right)^2 dr \ ,
		\end{equation}
		with $r_{\rm WS}$ being the WS-cell radius, $e$ denoting the elementary charge, and 
		\begin{equation}
			w(r) = \int_0^r (n_p(r')-n_e) r'^2 dr' \ .
		\end{equation}
		
		For the third term in Equation~\eqref{eq:ETF_total_free_energy_density}, the nuclear part $\mathcal{F}_{\rm nuc}$, we use the formulae up to the second-order Wigner--Kirkwood expansion in $\hbar$ available in Ref.~\cite{Onsi97}: 
		\begin{align}
			\mathcal{F}_{\text{nuc}} = \sum_q \left[ \frac{\hbar^2}{2m_q} 
			\left( \Gamma_{0q}^{(\text{TF})} + \Gamma_{0q}^{(2)}+\Lambda^{(2)}_{wq} 
			\right) \right] 
			+ \mathcal{V}_0 + \frac{Z_{\rm tot}}{V_{\rm WS}}m_pc^2 + \frac{N_{\rm tot}}{V_{\rm WS}}m_nc^2 \ ,
		\end{align}
		where $q=n,p$ for neutrons and protons, respectively, whose total numbers in the WS cell are denoted by $N_{\rm tot}$ and $Z_{\rm tot}$, while $c$ is the speed of light, and the potential part $\mathcal{V}_0$ is taken from the generalized Skyrme energy density functional BSk24~\cite{Goriely13} (see, e.g., Ref.~\cite{chamel2009} for explicit expression). 
		The Thomas--Fermi zero-order, second-order, and spin--orbit terms read, respectively,
		\begin{equation}
			\Gamma_{0q}^{(\text{TF})} = -\frac{1}{3\pi^2} \frac{1}{f_q^{3/2}} 
			\left( \frac{2m_q T}{\hbar^2} \right)^{5/2} I_{3/2}(\eta_q) 
			+ \frac{2m_q T}{\hbar^2} \eta_q n_q \ , 
		\end{equation}
		\begin{align}
			\Gamma_{0q}^{(2)} = \zeta_q f_q \frac{(\nabla n_q)^2}{n_q} 
			+ \left( \frac{9}{4} \zeta_q - \frac{7}{48} \right) n_q \frac{(\nabla f_q)^2}{f_q} 
			+ \frac{1}{6} (n_q \nabla^2 f_q - f_q \nabla^2 n_q)  \nonumber \\
			+ \left(3 \zeta_q - \frac{5}{12} \right) \nabla n_q \cdot \nabla f_q \ ,
		\end{align}
		\begin{equation}
			\Lambda^{(2)}_{wq}= -\frac{m^2_q \rho_q}{2\hbar^4 f_q}W^2_0\left[\nabla(n+n_q)\right]^2 \ ,
		\end{equation}
		with $f_q=m_q/m^\star_q$ representing the ratio between bare and effective masses, $W_0$ corresponding to the spin--orbit coupling, $n=n_p+n_n$, and%
		\begin{equation}
			\zeta_q = -\frac{I_{1/2}(\eta_q) \; I_{-(3/2)}(\eta_q)}{12 I_{-(1/2)}(\eta_q)^2} \ ,
		\end{equation}
		where we introduced the Fermi integrals\endnote{We compute the Fermi integrals of order 3/2 and 1/2 with the \texttt{GSL} routines available at 
			\url{https://gsl.ampl.com/ref/fermi-dirac.html} (accessed on 5th May 2025). For the Fermi integral of order $-$1/2 and $-$3/2 we rely on the formulae introduced in Ref.~\cite{Onsi94}.} of order $j$:
		\begin{align}
			I_j (\eta_q)  = \int_{0}^{\infty} \frac{t^j dt}{1+e^{t-\eta_q}} \ . 
			\label{eq:FD}
		\end{align}

		The function $\eta_q$ is obtained by inverting the relation:
		\begin{equation}
			n_q (r) = \frac{1}{2\pi}\left( \frac{2 m_q T}{\hbar^2 f_q} \right)^{3/2} I_{1/2} (\eta_q) \ .    
		\end{equation}

		The smoothly varying density distributions, $n_q(r)$, are parametrized with the soft-damping prescription \cite{Shchechilin24}
		\begin{equation}
			n_q (r) = n_{{\rm g}q}+n_{\Lambda q}f_q (r) \ ,   
		\end{equation}
		with
		\begin{equation}
			f_q (r) = \frac{1}{1+\left(\frac{C_q-r_{\rm WS}}{C_q}\right)^2\left(\frac{r}{r-r_{\rm WS}}\right)^2 e^{\frac{r-C_q}{a_q}}} \ .
			\label{eq:fq-etf}
		\end{equation}

		The parameters of the profiles $(n_{{\rm g}q}, n_{\Lambda q}, C_q, a_q)$ and the corresponding matter composition are obtained by minimizing the total free-energy density Equation~\eqref{eq:ETF_total_free_energy_density} at each given mean number density $n_B$ and temperature $T$. 
		As discussed previously, we set here $n_{{\rm g}p}=0$, therefore assuming that all protons are in the clustered part.  
		The number of protons and neutrons in the cell are then given by
		\begin{eqnarray}
			Z_{\rm tot}=Z&=& 4 \pi \, n_{\Lambda p}\int_{0}^{r_{\rm WS}}
			r^2 f_p(r)dr \ , \\
			N_\mathrm{tot}&=& 4 \pi \, n_{\Lambda n}\int_{0}^{r_{\rm WS}}
			r^2 f_n(r)dr + \frac{4\pi}{3}n_{{\rm g}n}r_{\rm WS}^3\ .
		\end{eqnarray}

		\subsection{The Temperature-Dependent CLDM}
		\label{sec:cldm}
		
		The treatment of the inner crust at finite temperature within the CLDM approach follows that of Ref.~\cite{dinh2023aa}; we recall here the main features.
		In the OCP approximation, the inner crust of an NS is considered to be composed of identical WS cells of volume $V_{\rm WS}$, each of which includes a fully ionized ion characterized by mass number $A$ and proton number $Z$, radius $r_N$, and internal density $n_i$, surrounded by uniform distributions of electrons and neutrons, with densities $n_e$ and $n_{\rm g}$, respectively.
		Moreover, as already mentioned, for consistency with ETF calculations performed in this work, we neglect the translational free-energy term that was included in Ref.~\cite{dinh2023aa}.
		
		At a given thermodynamic state, defined (if beta equilibrium is assumed) by the baryonic density $n_B$ and temperature $T$, the total free-energy density of the system can be written as
		\begin{equation}
			\mathcal{F} = \mathcal{F}_e +  \mathcal{F}_{\rm g} (1-u) + \frac{F_i}{V_{\rm WS}}, 
			\label{eq:total_free_energy_density}
		\end{equation}
		where $\mathcal{F}_e(n_e,T)$ is the free-energy density of the electron gas,
		$\mathcal{F}_{\rm g} = \mathcal{F}_B(n_{\rm g},\delta_{\rm g}, T) + m_n c^2 n_{\rm g}$ is the free-energy density of uniform neutron matter (including the rest masses of neutrons) at density $n_{\rm g}$ and isospin asymmetry $\delta_{\rm g} = 1$, $u=A/(n_i V_{\rm WS})$ is the volume fraction of the cluster, and $F_i$ is the cluster free energy, which includes the Coulomb interaction, as well as the residual interface interaction between the nucleus and the surrounding dilute nuclear-matter medium.
		The third term on the right-hand side of Equation~(\ref{eq:total_free_energy_density}), $-u \mathcal{F}_{\rm g} $, accounts for the excluded volume. 
		The total free-energy density $\mathcal{F}$ in Equation~(\ref{eq:total_free_energy_density}) is minimized under the constraint of baryon-number conservation:
		\begin{equation}
			n_B = n_{\rm g} + \frac{A}{V_{\rm WS}} \left(1 - \frac{n_{\rm g}}{n_i}\right) \ , 
			\label{eq:baryon_conservation}
		\end{equation}
		and charge neutrality holding, that is, $n_e = Z/V_{\rm WS}$. 
		
		\textls[-30]{The nuclear energetics is described employing a CLDM model approach, as in Refs.~\cite{Carreau2020a, dinh2021aa, dinh2023aa}. 
			Within this picture, the ion free energy, $F_i$ in Equation~(\ref{eq:total_free_energy_density}), can be written as:}
		\begin{equation}
			F_i = M_i c^2 + F_{\rm bulk} + F_{\rm Coul} + F_{\rm surf +  curv} \ ,
			\label{eq:Fi0}
		\end{equation}
		where $M_i = (A-Z)m_n + Zm_p$ is the total bare mass of the cluster, $F_{\rm bulk} = (A/n_i) \mathcal{F}_B(n_i, 1-2Z/A, T)$ is the cluster bulk free energy, and ${F}_{\rm Coul}$ and $F_{\rm surf +  curv}$ are the Coulomb and surface plus curvature energies, respectively.
		The bulk free energy is expressed in terms of the free-energy density of homogeneous nuclear matter, $\mathcal{F}_B(n,\delta,T)$, at a total baryonic density $n = n_n + n_p$ ($n_n$ and $n_p$ being the neutron and proton densities, respectively), isospin asymmetry $\delta = (n_n - n_p)/n$, and temperature $T$, following Refs.~\cite{lattimer1985, ducoin2007} (see Section~\ref{sec3.1} in Ref.~\cite{dinh2023aa} for details), where the free-energy density is decomposed into a `kinetic' and a `potential' part,
		\begin{equation}
			\mathcal{F}_B(n,\delta,T) = \mathcal{F}_{\rm kin}(n,\delta,T) + \mathcal{V}_{\rm MM}(n,\delta) \ .
			\label{eq:Fb}
		\end{equation}
		
		The `kinetic' term $\mathcal{F}_{\rm kin}$ in Equation~(\ref{eq:Fb}) is given by
		\begin{equation}
			\mathcal{F}_{\rm kin} = \sum_{q=n,p} \left[ \frac{-2 k_{\rm B} T}{\lambda_q^{3/2}} F_{3/2}\left( \frac{\tilde{\mu}_q}{k_{\rm B} T} \right) + n_q \tilde{\mu}_q  \right] \ ,
			\label{eq:Fkin}
		\end{equation}
		where $\lambda_q = \left(2 \pi \hbar^2/(k_{\rm B} T m^\star_q) \right)^{1/2}$ is the thermal wavelength of the nucleon with $m^\star_q$ being the density-dependent nucleon effective mass, 
		$F_{3/2}$ denotes the Fermi--Dirac integral, and the effective chemical potential $\tilde{\mu}_q$ is related to the thermodynamical chemical potential $\mu_q$ through $\tilde{\mu}_q = \mu_q - U_q$, with $U_q$ the mean-field potential (see Ref.~\cite{dinh2023aa} for details).
		Regarding the `potential' energy, we employ the meta-model (MM) approach of Ref.~\cite{Margueron2018a}, where $\mathcal{V}_{\rm MM}(n,\delta)$ is expressed as a Taylor expansion of order $N$ around the saturation point ($n=n_{\rm sat}$, $\delta=0$),
		\begin{equation}
			\mathcal{V}^{N}_{\rm MM}(n, \delta) = \sum_{k=0}^{N} \frac{n}{k!}(v^{\rm is}_{k} +v^{\rm iv}_{k}\delta^2 )x^{k}u^{N}_{k}(x),
			\label{eq:vMM}
		\end{equation}
		where $x = (n - n_{\rm sat})/(3n_{\rm sat})$, the parameters $v^{\rm is}_{k}$ and $v^{\rm iv}_{k}$ are linear combinations of the so-called nuclear-matter empirical parameters, and $u^{N}_{k}(x) = 1 - (-3x)^{N+1-k}\exp(-b(1+3x))$, with $b = 10\ln(2)$ being a parameter governing the functional behavior at low densities (see Ref.~\cite{Margueron2018a} for details). 
		The authors of Ref.~\cite{Margueron2018a} showed that different nucleonic energy functionals at zero temperature can be satisfactorily reproduced by truncating the expansion, Equation~\eqref{eq:vMM}, at order $N=4$. 
		The MM parameter set thus has 13 parameters, namely, the nuclear-matter empirical parameters (up to order $N=4$), together with $n_{\rm sat}$, the isoscalar effective mass, the effective mass splitting, and the parameter $b$. The homogeneous-matter energy of a given functional at subsaturation can, therefore, be reproduced by assigning the bulk nuclear-matter parameters corresponding to those of the functional. In this work, we employ the parameter set associated with the BSk24 functional \cite{Goriely13}, for consistency with the ETF calculations.
		We note that the same expression of the bulk energy is used to compute the free-energy density of the free neutron gas, $\mathcal{F}_{\rm g}$, thus ensuring a consistent treatment of nucleons. 
		
		The Coulomb and finite-size contributions are written as $F_{\rm Coul} + F_{\rm surf +  curv} = V_{\rm WS} (\mathcal{F}_{\rm Coul} + \mathcal{F}_{\rm surf + curv})$. 
		As we consider here only spherical clusters, the Coulomb energy density is given by \cite{Ravenhall83}
		\begin{equation}
			\mathcal{F}_{\rm Coul}  = \frac{2}{5}\pi (e n_i r_N)^2 u \left(\frac{1-\delta_{\rm cl}}{2}\right)^2 \left[ u+ 2  \left( 1- \frac{3}{2}u^{1/3} \right) \right] \ , 
			\label{eq:Fcoul}
		\end{equation}
		with $\delta_{\rm cl} = 1-2Z/A$.
		For the surface and curvature contributions, we employ the same expressions as in Refs.~\cite{Lattimer91, Maruyama2005, Newton2013}, that is
		\begin{equation}
			\mathcal{F}_{\rm {surf + curv}} =\frac{3u}{r_N} \left( \sigma_{\rm s}(\delta_{\rm cl}, T) +\frac{2\sigma_{\rm c}(\delta_{\rm cl}, T)}{r_N} \right) \ , 
			\label{eq:interface}   
		\end{equation}
		where $\sigma_{\rm s}$ and $\sigma_{\rm c}$ are the surface and curvature tensions \cite{Lattimer91},
		\begin{equation}
			\sigma_{\rm s(c)}(\delta_{\rm cl},T) = \sigma_{\rm s(c)}(\delta_{\rm cl}, T=0) h(T) \ .
			\label{eq:sigmaT}
		\end{equation}
		
		The expressions of the surface and curvature tensions at zero temperature are taken from Refs.~\cite{Ravenhall1983, Maruyama2005, Newton2013}, who proposed a parameterization based on Thomas--Fermi calculations at extreme isospin asymmetries,
		\begin{eqnarray}
			\sigma_{\rm s}(\delta_{\rm cl}, T=0) &=& \sigma_0 \frac{2^{p+1} + b_{\rm s}}{y_p^{-p} + b_{\rm s} + (1-y_p)^{-p}} \ , \label{eq:sigmas} \\
			\sigma_{\rm c} (\delta_{\rm cl},T=0) &=& 5.5 \sigma_{\rm s}(\delta_{\rm cl},T=0) \frac{\sigma_{0, {\rm c}}}{\sigma_0} (\beta -y_p) \ ,
			\label{eq:sigma0}
		\end{eqnarray}
		where $y_p = (1-\delta_{\rm cl})/2$.
		The temperature dependence of the surface terms is given by \citep{Lattimer91}
		\begin{equation}
			h(T) = 
			\begin{cases} 
				0 & \text{if $T > T_{\rm c}$}  \\
				\left[1 -\left(\frac{T}{T_{\rm c}}\right)^2\right]^2 & \text{if $T \leq T_{\rm c}$}
			\end{cases} \ ,
			\label{eq:hT}
		\end{equation}
		where $T_{\rm c}$ is the critical temperature (see Equation~(2.31) in Ref.~\cite{Lattimer91}).
		However, to compare with ETF calculations in this work, we use $\sigma_{\rm s(c)} = \sigma_{\rm s(c)}(\delta_{\rm cl}, T=0)$ (that is, we set $h(T)=1$ at all temperatures\endnote{We checked that for all the conditions explored in this work, the temperature always remains below the critical temperature.}).
		This is because the temperature dependence encoded in \mbox{Equations~\eqref{eq:sigmaT} and \eqref{eq:hT}} is supposed to come from the scaling associated with the behavior of the surface tension near the critical temperature, which depends on beyond-mean-field effects not included in the approaches employed in this work. 
		A precise fit from the ETF calculations of the isospin-dependent critical temperature, and the associated temperature dependence of the surface energy, is beyond the scope of the present paper. 
		The surface parameters in Equations~\eqref{eq:sigmas} and \eqref{eq:sigma0}, namely $(\sigma_0, \sigma_{0, {\rm c}}, b_{\rm s}, \beta,p)$, are optimized for the set of bulk parameters and effective mass corresponding to the BSk24 functional in three different~ways:
		\begin{enumerate}
			\item[(i)] A fit of the surface energy, $E_{\rm surf+curv} = V_{\rm WS} \mathcal{F}_{\rm surf+curv}(\delta_{\rm cl},T=0)$ (see Equation~\eqref{eq:e-etf-cldm}), parametrized as Equation~\eqref{eq:interface}, as extracted from ETF calculations in the medium\endnote{For this fit, the ETF energy is calculated for different fixed proton fractions and ion proton numbers, associated with different background gas densities; see Appendix~\ref{app:etf-fit} for details.}
			at $T=0$ \cite{pearson2018,pearson_priv} performed using the BSk24 functional. The followed procedure is similar to that described in Ref.~\cite{Furtado2021} (see Appendix~\ref{app:etf-fit} for details).
			\item[(ii)] A fit of nuclear masses in vacuum computed with the CLDM,
			\begin{eqnarray}
				M_{\rm nuc}(A,Z) c^2 &=& m_p c^2 Z+m_n c^2 (A-Z) + \frac{A}{n_i} \mathcal{E}_B(n_i,\delta_{\rm cl}) \nonumber \\
				&+& 4\pi r_N^2\left (\sigma_s +\frac{2\sigma_c}{r_N}\right ) + \frac{3}{5}\frac{e^2 Z^2}{r_N} \ , 
				\label{eq:mass-cldm}
			\end{eqnarray}
			where $\mathcal{E}_B$ is the bulk energy density, to the ETF mass table for all nuclei up to nuclear drip lines, from $Z=8$ up to $Z=122$, obtained using the BSk24 functional\endnote{The ETF mass table calculated with the BSk24 functional is available at 
				\url{https://cdsarc.cds.unistra.fr/viz-bin/cat/J/A+A/635/A84} (accessed on 5th May 2025).}, as was done in Ref.~\cite{Carreau2020a} (see also their Table~1).
			\item[(iii)] \textls[-25]{A fit of the nuclear masses, Equation~\eqref{eq:mass-cldm}, to reproduce the
				experimental atomic masses in the 2020 Atomic Mass Evaluation (AME) table \cite{AME2020}, as was performed in previous works \cite{dinh2023aa}\endnote{In Ref.~\cite{dinh2023aa} the fit of the surface parameters was performed to reproduce the AME2016 table \cite{AME2016}.}}.
		\end{enumerate}
		
		In cases (ii) and (iii), the parameter $p$ in Equations~\eqref{eq:sigmas} and \eqref{eq:sigma0} governing the behavior of the surface tension at high isospin is set to $p=3$ as this value was seen to provide a good reproduction of the crust-core transition point obtained with the BSk24 model in Ref.~\cite{pearson2018}; we note that this optimal value is also confirmed by the fit (i) (see Table~\ref{tab:bestfit_bsk24}).
		Comparing with the results of Refs.~\cite{Furtado2021,balliet2021}, we obtain a slightly lower value for the best-fit value of the $p$ parameter, $p=3$ instead of $p=3.4$--
		$3.9$. 
		This apparent discrepancy might be explained by the differences in the fitting procedure, namely, we employ a different functional (BSk24 instead of SLy4), the expression for the curvature tension we use (see Equation~\eqref{eq:sigma0}) has an additional $(\beta - y_p)$ dependence with respect to Refs.~\cite{Furtado2021,balliet2021} (see Equation~(49) in Ref.~\cite{Furtado2021}), the ETF code we employ for the fit differs from that of Ref.~\cite{Furtado2021} (for instance, we implement a damped Fermi profile instead of a standard Fermi profile for the nucleon densities; see also Appendix~\ref{app:etf-fit}), and the pool of nuclei considered for the fit is not the same\endnote{We actually checked the impact of the above-mentioned points and fitted the surface parameters on ETF calculations in the medium using a standard Fermi profile and the SLy4 functional, and including the proton exchange in the Coulomb term. The best-fit value of the $p$ parameter thus obtained is $p=3.5$ if we use the parametrization of the curvature tension as in Equation~\eqref{eq:sigma0}, and $p=3.6$ if we use Equation~(49) in Ref.~\cite{Furtado2021}, that is, neglecting the $(\beta - y_p)$ term. These latter results are in better agreement with those of Refs.~\cite{Furtado2021,balliet2021}, confirming that the differences in the fitting procedure mentioned above can account for the discrepancy in the resulting $p$ values.}.
		The results of the best fits we obtain for the surface-energy parameters are given in Table~\ref{tab:bestfit_bsk24}.
		
		\begin{table}[H] 
			\caption{Best 
				fit to ETF calculations in the medium at zero temperature (first row), full ETF mass table (second row), and to the AME2020 table (third row) for the surface parameters in the CLDM for the BSk24 functional. 
			}
			\begin{tabularx}{\textwidth}{lccccc}
				\toprule
				& $\sigma_0$ & $\sigma_{0,{\rm c}}$ & $b_{\rm s}$ & $\beta$ & $p$ \\
				\midrule
				ETF calculations in the medium
				& 1.033329 & 0.172145 & 29.418725 & 0.747555 & 3.0 \\
				ETF mass table & 0.98636  & 0.09008 & 36.22714 & 1.16310 & 3.0 \\
				AME2020        & 1.04971  & 0.12094 & 30.45764 & 0.66720 & 3.0 \\
				\bottomrule
				\label{tab:bestfit_bsk24}
			\end{tabularx}
		\end{table}
		\vspace{-22pt}
		\section{Neutron Star Crust at Finite Temperature}\label{sec:results}
		
		\subsection{Thermodynamic Properties and Composition}\label{sec3.1}
		
\begin{figure}[H]
			\includegraphics[width=10.cm]{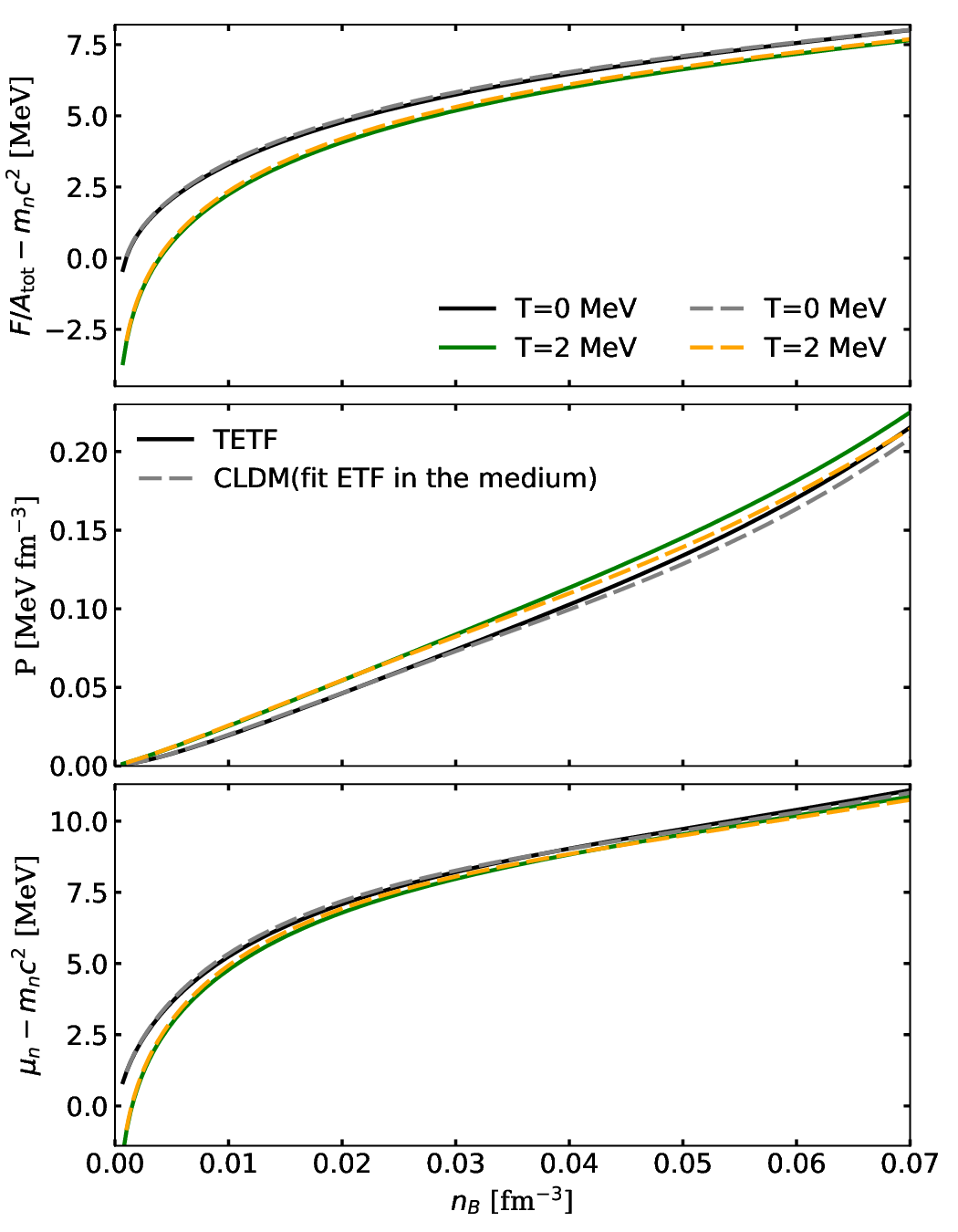}
			\caption{Free energy per nucleon (with the neutron mass subtracted out, top panel), pressure (middle panel), and neutron chemical potential (bottom panel) in the WS cell as a function of the baryon density for $T=2$~MeV. Results are obtained with the TETF (solid lines) and CLDM with surface parameters fitted to ETF calculations in the medium (dashed lines). 
				For comparison, results at $T=0$~MeV are also shown. }
			\label{fig:thermo}
		\end{figure}

		In this section, we analyze the effects of temperature on the properties of the NS inner crust in beta equilibrium.
		We start by showing in Figure~\ref{fig:thermo} the thermodynamic properties obtained with the TETF (solid lines) and with the CLDM with surface parameters fitted to ETF calculations in the medium (dashed lines) (all data used to produce the figures of this work are avaible as supplementary material to this article).
		In all calculations, we employ the BSk24 functional, as described in Section~\ref{sec:method}.
		The top panel shows the total free energy per nucleon in the WS cell (with the neutron mass subtracted out), while the middle (bottom) panel displays the pressure (neutron chemical potential) in the WS cell as a function of the baryon number density in the inner crust, at the selected temperature of $T=2$~MeV. For comparison, results at $T=0$ MeV are also shown.
		From the top panel, we note a decrease of $F/A_{\rm tot}$ as the temperature increases, with a higher impact at lower densities, while from the middle panel we see an increase of pressure with temperature. 
		We note that the CLDM results almost overlap with the TETF ones, with the CLDM predicting slightly lower pressures for $n_B \gtrsim 0.03$ fm$^{-3}$.
		As for the neutron chemical potential, shown in the bottom panel, we note that at higher densities, the numerical results at the different considered temperatures predict almost identical values due to the degeneracy of the neutron gas, which is, however, lifted by the finite-temperature effect at very low densities leading to the noticeable smaller $\mu_n$ at $T=2$~MeV. 
		In summary, Figure~\ref{fig:thermo} shows that the predictions of the CLDM are in good agreement with TETF calculations as far as the thermodynamic properties of the NS inner crust are concerned.

\begin{figure}[H]
			\includegraphics[width=10. cm]{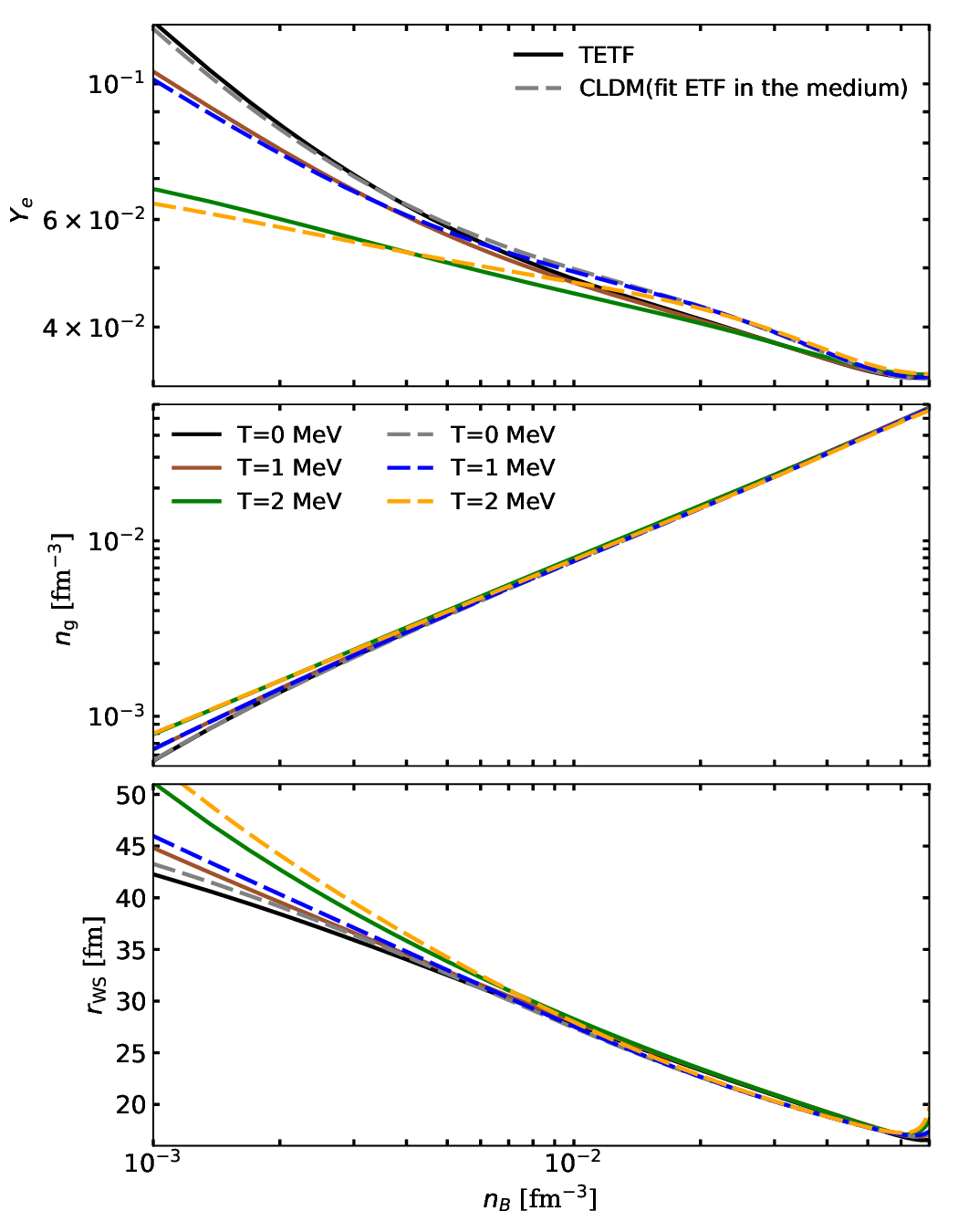}
			\caption{Electron fraction (top panel), neutron-gas density (middle panel), and WS-cell radius (bottom panel) as a function of the baryon density for $T=1$~MeV and $2$~MeV. Results are obtained with the TETF (solid lines) and CLDM with surface parameters fitted to ETF calculations in the medium (dashed lines). For comparison, results at $T=0$~MeV are also shown.
			}
			\label{fig:crust}
		\end{figure}

		Additional properties of the inner crust are shown in Figure~\ref{fig:crust}, where we display the electron fraction $Y_e = n_e/n_B$ (top panel), the background neutron-gas density (middle panel), and the WS-cell radius (bottom panel). 
		We note that the temperature impacts $Y_e$ mainly at low densities, with higher temperature yielding lower electron fraction.
		Moreover, for all temperatures, a constant decrease of $Y_e$ is observed as the density increases. 
		For $n_B \gtrsim 0.02$ fm$^{-3}$, the results obtained within the chosen model, BSk24, are almost identical for the different selected temperatures. 
		We also note that at low densities, TETF predictions for the electron fraction are usually higher than CLDM ones.
		The neutron-gas density shows almost no change with temperature, particularly as density increases, with higher $T$ producing higher $n_{\rm g}$ up to $n_B \approx 3 \times 10^{-3}$ fm$^{-3}$. 
		On the other hand, a non-negligible effect of the temperature is observed on the WS-cell radius at lower densities, as can be seen from the bottom panel, with higher temperature producing larger $r_{\rm WS}$; however, for \mbox{$n_B \gtrsim 0.01$ fm$^{-3}$}, the impact of temperature is negligible.
		We show in Figures~\ref{fig:thermo} and \ref{fig:crust} the results of the CLDM with surface parameters fitted on ETF calculations in the medium as an illustrative example. 
		Indeed, the curves obtained using the CLDM with surface parameters fitted as in the protocols (ii) and (iii) (see Section~\ref{sec:cldm}) are hardly distinguishable.
		
\begin{figure}[H]
			\includegraphics[width=14. cm]{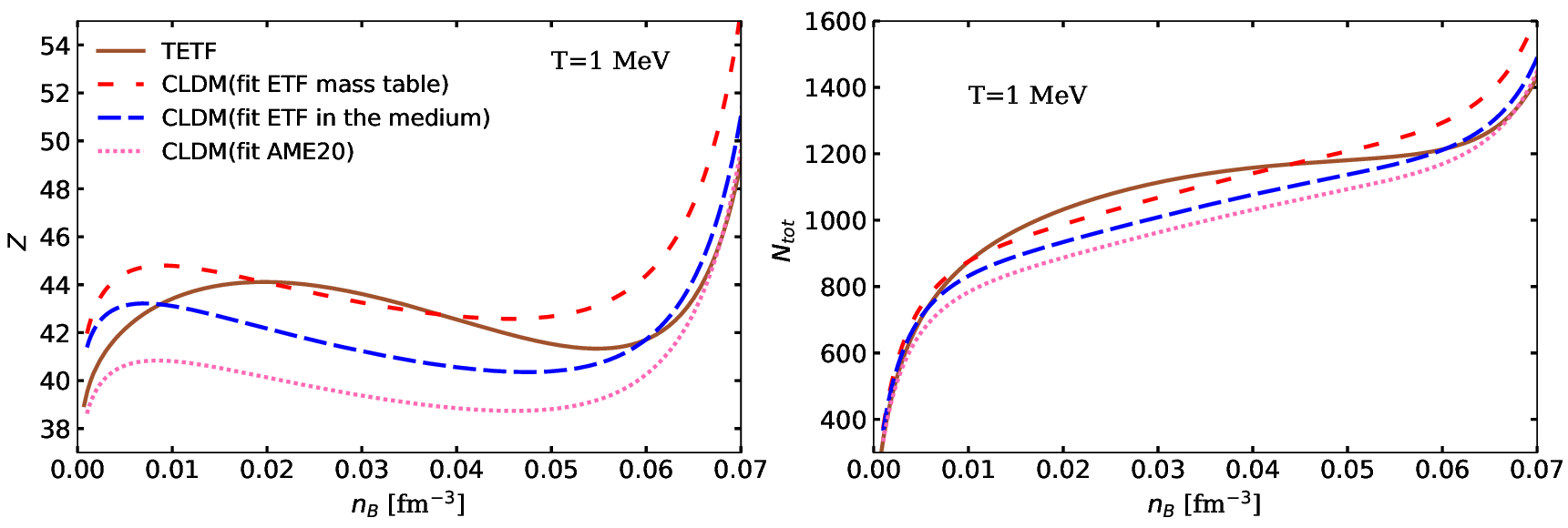}\\
			\includegraphics[width=14. cm]{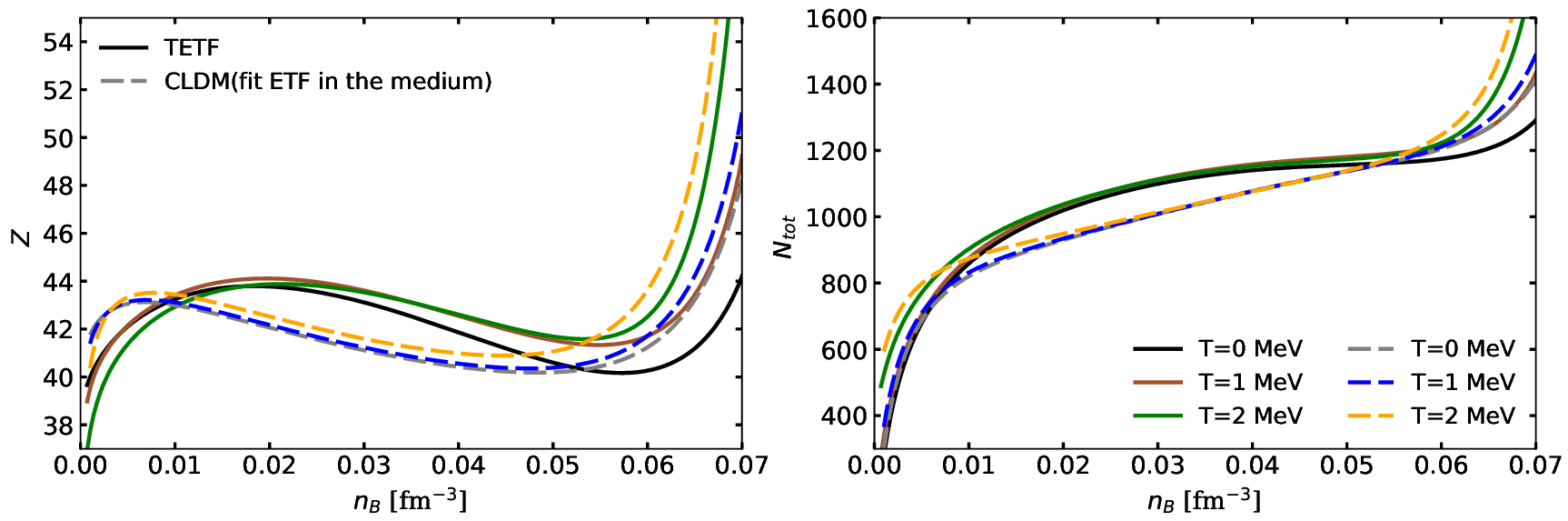}
			\caption{Number of protons (left panels) and total number of neutrons (right panels) in the WS cell as a function of the baryon density. Top panel: results are shown for $T=1$~MeV for TETF (solid lines) and CLDM with surface parameters fitted to full ETF mass table (red dashed lines), to ETF calculations in the medium at $T=0$~MeV (blue dashed lines), and to the AME2020 table (pink dotted lines).
				Bottom panel: results are shown for $T=1,2$~MeV for TETF (solid lines) and CLDM with surface parameters fitted to ETF calculations in the medium (dashed lines). For comparison, results at $T=0$~MeV are also shown. 
			}
			\label{fig:ZclNcl}
		\end{figure}

		We turn now to the discussion of the crust composition, that is, the proton and total neutron number in the WS cell, shown in the left and right panels, respectively, of Figure~\ref{fig:ZclNcl}.
		To analyze the impact of the surface-fitting protocol in the CLDM, we display in the top panels the results obtained with the CLDM with surface parameters fitted to ETF calculations in the medium at $T=0$ (fit protocol (i), blue dashed lines), to ETF nuclear mass table (fit protocol (ii), red dashed lines), and to the AME2020 table (fit protocol (iii), pink dotted lines), at the selected temperature of $T=1$~MeV.
		Results obtained with TETF are shown as solid lines.
		The qualitative behavior of $Z$ and $N_{\rm tot}$ is the same in the different calculations.
		We note that the CLDM with surface parameters fitted to the ETF mass tables and to the ETF calculations in the medium (see also the discussion in Appendix~\ref{sec:surf-en} and Figure~\ref{fig:Es}) are in slightly better agreement with the TETF results than those obtained with the CLDM surface parameters fine-tuned to the experimental AME2020 table.
		To analyze the impact of temperature on the composition, we show the results obtained with TETF (solid lines) and CLDM with surface parameters adjusted to ETF calculations in the medium (dashed lines) in the bottom panels of Figure~\ref{fig:ZclNcl}, for the selected temperatures of $T=1$~MeV and $T=2$~MeV. For comparison, results at $T=0$~MeV are also shown.
		Although the considered CLDM underestimates the nucleon numbers in the interval \mbox{$0.01$ fm$^{-3}$ $\lesssim n_B \lesssim$ $0.05$ fm$^{-3}$} and overestimates them as the density approaches the crust-core transition, the trend followed is the same at all considered temperatures.
		One can see that both the neutron and proton numbers grow with temperature at high densities; on the other hand, the proton number lowers for higher temperatures at low densities.
		
		\vspace{-12pt}

		\subsection{Density Profiles at Zero and Finite Temperature}
		
\begin{figure}[H]
			\includegraphics[width=13.6 cm]{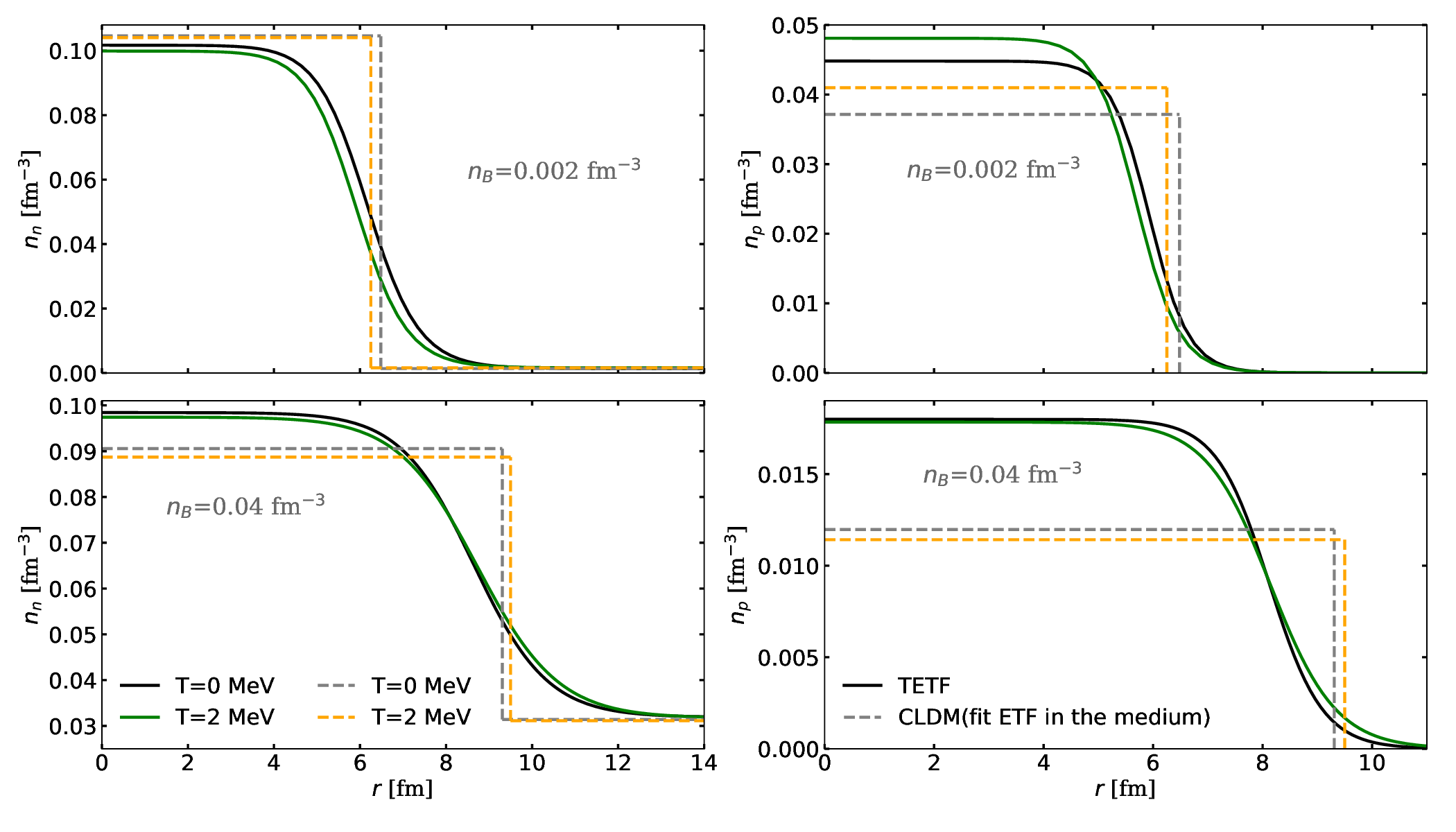}
			\caption{Neutron (left panels) and proton (right panels) density profiles at two selected baryon densities in the NS inner crust: $n_B = 0.002$~fm$^{-3}$ (top panels) and $n_B=0.04$~fm$^{-3}$ (bottom panels), and at the selected temperature of $T=2$~MeV. 
				Solid (dashed) lines show predictions for TETF (CLDM with surface parameters fitted to ETF calculations in the medium). 
				For comparison, results at $T=0$~MeV are also shown.
			}
			\label{fig:profiles}
		\end{figure}

		As mentioned in the introduction, nucleon density profiles are important to calculate, for example, the electron conductivity coefficients in the crust, since they depend on the charge distribution \cite{Kaminker1999,Gnedin2001}.
		We show in Figure~\ref{fig:profiles} the density profiles for neutrons (left panels) and protons (right panels) in the NS crust obtained with the TETF (solid lines), for $T=0, 2$~MeV and two selected densities: $n_B = 0.002$~fm$^{-3}$ (top panels) and $n_B = 0.04$~fm$^{-3}$ (bottom panels).
		We also display the density profiles extracted from the CLDM with surface parameters fitted to ETF calculations in the medium (dashed lines).
		To this aim, we use the cluster radius $r_N$ and the total cluster density $n_i$ as determined from the minimization procedure; we then defined the central cluster density $n_{Np}$ from the charge conservation, Equation~\eqref{eq:Z-bar-cons}, and the neutron cluster density as $n_{Nn} = n_i - n_{Np}$. 
		We note that at lower density (top left panel) the neutron density at the center of the cell is slightly smaller at finite temperature, and the neutron cluster density from the CLDM well reproduces the TETF results.
		For higher density (bottom left panel) the CLDM predicts lower values for the neutron cluster density when compared to TETF.
		As for the proton densities, at lower density (top right panel), we note an increase of the proton central density at finite temperatures, with CLDM following the same trend but predicting lower values with respect to the TETF.
		At higher density (bottom right panel), the TETF proton central density is basically unchanged with temperature, while the CLDM predicts lower values of the cluster proton density with respect to the TETF and yields slightly smaller values for higher temperatures.
		The impact of baryon density on the profiles is pronounced, amplifying the differences between CLDM and TETF.  
		Additionally, at higher densities, the expected effect of temperature becomes more apparent, with increasing temperature leading to broader TETF profiles compared to the zero-temperature case.
		It has to be noted that the differences observed between the two approaches, TETF and CLDM, trace back to the definition of the radius (only one radius is defined in the CLDM, as explained in Appendix~\ref{app:etf-fit}), leading to the overestimation of the proton radii in the CLDM case.
		Moreover, the equilibrium compositions at the shown densities are different in the CLDM and TETF (see Figure~\ref{fig:ZclNcl}).

		\section{Conclusions}
		\label{sec:concl}
		
		In this work, we have investigated the effects of temperature on the inner crust of a non-accreting NS by comparing calculations from the semi-classical TETF approach with the CLDM. 
		In both approaches, the same functional (BSk24) was employed.
		Our findings confirm that thermodynamic properties such as energy density, pressure, and chemical potentials are reasonably well reproduced by the CLDM, with minor differences compared to TETF predictions. 
		Small deviations also arise in certain crust properties, such as the electron fraction and the WS-cell radius, particularly at low densities.
		
		We show that the considered CLDM can grasp the behavior of the TETF predictions concerning the NS inner-crust composition, including the changes of nucleon numbers with temperature. Namely, both methods predict the proton number to grow (decrease) with temperature at higher (lower) mean baryon densities.
		However, some quantitative discrepancies between the predictions of TETF and CLDM are observed. This arises from the different treatments of finite-size effects, as was previously found at zero-temperature~\cite{Grams_2022}.
		In particular, the CLDM with surface parameters fitted to the experimental atomic mass tables generally yields lower $Z$ and $N_{\rm tot}$ than the ones found in the TETF approach. 
		Our results suggest that the CLDM reproduction of the TETF crustal composition can be improved if information on extremely isospin asymmetric matter (that is absent in the experimental data) is included in the fitting procedure. 
		This is incorporated in our CLDM when the surface-energy parameters are fine tuned to the ETF calculations in the medium at $T=0$~MeV or to the full ETF mass tables (from proton to neutron drip line). 
		This observation is in agreement with the findings of Ref.~\cite{Carreau2020a}, where it was highlighted that accounting for the surface properties at extreme isospin values is important for the determination of the crystallization temperature and the associated inner-crust composition.
		Nevertheless, the overall agreement with TETF results is achieved even with the CLDM fitted to the experimental data only, suggesting that this would be a tool of choice for very fast computations required in statistical analyses.
		
		On the other hand, nucleon density profiles can only be well determined in the TETF approach, especially at high densities, where the nucleon density distributions broaden and difference between neutron and proton densities increases. Moreover, the absence of the neutron skin in the CLDM can lead to too large proton radii as compared to the TETF predictions. 
		The TETF is, thus, a preferable choice in applications where an accurate description of the microphysics is needed.
		This method can be further improved by consistently incorporating quantum shell corrections and pairing correlations at low temperatures with the ETFSI~\cite{Chamel2024}. 
		Such an approach can be considered as a computationally fast approximation of the full HFB equations \cite{Shelley2020}. 
		However, the latter may be required for probing the formation of nuclear pasta in the deepest layers of the crust~\cite{Shchechilin24}. Full HFB calculations are also unavoidable for studying the nuclear dynamics~\cite{Pecak2024}.
		
		The comparisons of crustal properties between CLDM and TETF method could be extended by applying other Skyrme functionals that differ, for example, in the symmetry energy behavior. This would allow to investigate the role of bulk properties in the crust composition and EoS (see, e.g., Refs.~\cite{dinh2021aa, dinh2023aa} for such a study within zero- and finite-temperature CLDM and Ref.~\cite{pearson2018} within ETF). However, we expect that the change of functional would not alter the key conclusions of this work, in particular, the observed discrepancies between the CLDM and TETF approach related to the cluster composition, density profiles, and the role of the neutron skin.
		
		The results presented here are particularly relevant in the current era of gravitational-wave and multi-messenger astronomy, where accurate modeling of the NS---including the crust and its finite-temperature behavior---is essential for interpreting observations from binary NS mergers and extract dense-matter properties. 
		As current EoS tables used in simulations rely on simplified models, our study offers a necessary benchmark, clarifying where computationally fast approaches like CLDM suffice and where more detailed treatments like TETF become indispensable for unified EoS construction and crustal microphysics. 
		In particular, we show that reliable calculations of transport properties require accurate density profiles, which can indeed be obtained from TETF.

		\vspace{6pt} 
		
		
		\supplementary{The following supporting information can be downloaded at: \linksupplementary{s1}. The dataset includes numerical results used to generate the figures in the article. 
		A detailed README file is included to describe file structure and column content.}

		
		\authorcontributions{Conceptualization
			, all authors; methodology, all authors; software, G.G., N.N.S., A.F.F., and T.D.; validation, all authors; formal analysis, G.G., N.N.S., T.D., and A.F.F.; investigation, all authors; resources, N.C., A.F.F., and F.G.; data curation, G.G., N.N.S., T.D., and A.F.F.; writing—original draft preparation, G.G., and A.F.F.; writing—review and editing, all authors; visualization, G.G. and T.D.
			All authors have read and agreed to the published version of the manuscript.}
		
		\funding{The work of N.N.S. and G.G. was financially supported by the FWO (Belgium) and the Fonds de la Recherche Scientifique (Belgium) under the Excellence of Science (EOS) program (project No. 40007501). The work of N.C. received funding from the Fonds de la Recherche Scientifique (Belgium) under Grant Number IISN 4.4502.19. This work was also partially supported by the CNRS International Research Project (IRP) ``Origine des \'el\'ements lourds dans l’univers: Astres Compacts et Nucl\'eosynth\`ese (ACNu)''. }

		\dataavailability{Data are contained within the article
			.}

		\acknowledgments{G.G. and N.N.S. thank Hoa Dinh 
			Thi for insightful discussions on the temperature dependence of nuclear matter in neutron star environments, which greatly contributed to this~work.}
		
		\conflictsofinterest{The authors declare no conflicts of interest.}

		
		
		\abbreviations{Abbreviations}{
			The following abbreviations are used in this manuscript:\\
			
			\noindent 
			\begin{tabular}{@{}ll}
				AME & Atomic Mass Evaluation \\
				CLDM & compressible liquid-drop model \\
				EoS & equation of state \\
				(E)TF & (extended) Thomas--Fermi\\
				ETFSI & extended Thomas--Fermi plus Strutinsky integral \\
				HFB & Hartree--Fock--Bogoliubov \\
				MM & meta-model \\
				NS & neutron star \\
				OCP & one-component plasma \\
				TTF & temperature-dependent Thomas--Fermi\\
				TETF & temperature-dependent extended Thomas--Fermi\\
				TETFSI & temperature-dependent extended Thomas--Fermi plus Strutinsky integral\\
				WS & Wigner--Seitz 
			\end{tabular}
		}

		\appendixtitles{yes} 
		\appendixstart
		\appendix
		
		\section{Fit of the CLDM Surface Parameters to ETF Calculations}
		\label{app:etf-fit}
		
		In the CLDM employed in this work, for a given set of nuclear-matter empirical (bulk) parameters, the surface parameters are determined consistently through a fit.
		In the present study, the bulk parameters are fixed to those corresponding to BSk24, and the surface parameters are fitted in three different ways, as mentioned in Section~\ref{sec:cldm}: 
		(i) from a fit to ETF calculations of clusters in the medium at $T=0$ \cite{pearson2018, pearson_priv}, 
		(ii) from a fit to the ETF mass table calculated with the BSk24 functional (see Ref.~\cite{Carreau2020a} for details), 
		and (iii) from a fit to the AME2020 \cite{AME2020} experimental mass table (see, e.g., Ref.~\cite{dinh2021aa} for details). 
		We describe here the procedure adopted for the fit (i), which was performed along the line of Ref.~\cite{Furtado2021}.
		To extract the surface energy from ETF calculations at $T=0$, we decompose the WS-cell energy (with the electron-gas energy and the constituent masses subtracted out) as 
		\begin{equation}
			E_{\rm ETF} = \mathcal{E}_B(n_{Np},n_{Nn}) V_N + E_{\rm surf+curv} + E_{\rm Coul} + \mathcal{E}_B(0,n_{\rm g}) (V_{\rm WS} - V_N) \ ,
			\label{eq:e-etf-cldm}
		\end{equation}
		where $V_N$ is the cluster volume, $n_{Nn}$ ($n_{Np}$) is the density of neutrons (protons) in the cluster, and $n_{\rm g}$ is the gas density.
		As the calculations were performed at zero temperature, only neutrons are considered in the gas; thus, $n_{\rm g} = n_{{\rm g}n}$ and $n_{{\rm g}p}=0$.
		The bulk energy densities in Equation~\eqref{eq:e-etf-cldm}, $\mathcal{E}_B$, both in the cluster and in the nucleon gas, correspond to the energy densities of a homogeneous system at the corresponding neutron and proton densities and can be expressed in terms of the Skyrme parameters corresponding to the BSk24 functional, see Refs.~\cite{pearson2018, chamel2009} for details and complete expressions.
		The Coulomb energy in Equation~\eqref{eq:e-etf-cldm} is calculated from the Gaussian theorem assuming a constant density profile in the cluster and in the background gas, and reads:
		\begin{equation}
			E_{\rm Coul} = \frac{3}{5} \frac{(eZ)^2}{r_N} \left( 1-\frac{3}{2}\frac{r_N}{r_{\rm WS}} + \frac{1}{2}\frac{r_N^3}{r_{\rm WS}^3}\right) 
			+ \frac{3 e^2 Z}{8}\left( \frac{3}{\pi}\right)^{1/3} n_e^{1/3} \ ,
			\label{eq:e-coul-cldm}
		\end{equation}
		where the last term accounts for the electron-exchange contribution, that is included in Ref.~\cite{pearson2018} (see Equation~(B10) in Ref.~\cite{pearson2018}).
		As discussed in Ref.~\cite{Furtado2021}, this means that the surface energy in Equation~\eqref{eq:e-etf-cldm} contains a residual Coulomb contribution coming from the presence of an interface between the nucleus and the background gas.
		
		\textls[-15]{In order to extract $E_{\rm surf+curv}$, which we parametrize as $E_{\rm surf+curv} = V_{\rm WS} \mathcal{F}_{\rm surf+curv}(\delta_{\rm cl}, \mbox{T=0})$ according to Ref.~\cite{Lattimer91} (see Equations~\eqref{eq:interface}--\eqref{eq:sigma0}), one needs to specify the cluster density and radius.
			In this work, we employ the simplest prescription ignoring the neutron skin\endnote{We have also performed a fit including a neutron skin. However, this yielded a higher $\chi^2$, in accordance with the results of Ref.~\cite{Furtado2021}. Therefore, we only kept the prescription without the neutron skin for this work.}, and thus, the cluster is defined by one radius only, $r_N$, and $V_N = 4 \pi r_N^3/3$.
			For the cluster density, different choices are possible, as discussed in Ref.~\cite{Furtado2021}.
			We choose to define the cluster total density as the total central density extracted from ETF calculations,}
		\begin{equation}
			n_i = \sum_q n_{Nq} = \sum_q n_q(r=0) \ , 
			\label{eq:ncl-etf}
		\end{equation}
		where $n_q(r=0)$ is the nucleon density at the center of the WS cell.
		We note that the parametrized density profile in the ETF calculations used for the fitting procedure (see Equations~(14) and (15) in Ref.~\cite{pearson2018}) slightly differs from that employed in the TETF calculations performed here, Equation~\eqref{eq:fq-etf}.
		From the total cluster density $n_i$, the cluster radius $r_N$ is extracted from the baryon number conservation equation in the WS cell,
		\begin{equation}
			A_{\rm tot} = \frac{4}{3} \pi \left[ n_i r_N^3 + n_g \left(r_{\rm WS}^3 - r_N^3 \right) \right] \ ,
			\label{eq:A-bar-cons}
		\end{equation}
		the cluster proton density is then obtained from the charge conservation,
		\begin{equation}
			Z = \frac{4}{3} \pi n_{Np} r_N^3 \ ,
			\label{eq:Z-bar-cons}
		\end{equation}
		and finally, the cluster neutron density is simply given by $n_{Nn} = n_i - n_{Np}$.
		We note that $n_{Nq}$ is different from $n_q(r=0)$ because in the CLDM only one radius is defined.
		
		The fit of the surface parameters was performed with the \texttt{scipy.curvfit} routine, using 334 clusters in the medium.
		The ETF calculations spanned a baryon-density range from $\sim$$5 \times 10^{-4}$~fm$^{-3}$ to $\sim$$ 0.02$~fm$^{-3}$ and were carried out either at fixed total proton fraction or at fixed $Z$ in order to have in the fitting pool a large cluster-asymmetry range (the total proton fraction ranged approximately from 0.04 to 0.5 corresponding to a cluster proton fraction approximately from 0.15 to 0.5).

		\section{Comparison of the Surface Energy Between the TETF and the CLDM}
		\label{sec:surf-en}
		
		To extract the surface energy in the TETF in order to compare with the CLDM, we define it as:
		\begin{equation} 
			F_{\rm surf}^{\rm TETF} = F_{\rm nuc}^{\rm TETF} - V_N \mathcal{F}_B(n_{Np},n_{Nn}) -  (V_{\rm WS}-V_N) \mathcal{F}_B(0,n_{{\rm g}n}) \ ,
			\label{eq:fsurf-etf}
		\end{equation}
		where $V_N$, $r_N, n_{Np}$ and $n_{Nn}$ 
		are extracted from Equations~\eqref{eq:A-bar-cons} and \eqref{eq:Z-bar-cons} of Appendix~\ref{app:etf-fit}, and $F_{\rm nuc}^{\rm TETF}$ is the nuclear-matter energy of the WS cell obtained with the TETF approach, after subtracting out the electron and Coulomb terms from Equation~\eqref{eq:ETF_total_free_energy_density}.
		Note, however, that as in Appendix~\ref{app:etf-fit}, $n_{Np}$ and $n_{Nn}$ do not coincide with the neutron and proton central densities within the TETF method. This is due to the absence of the neutron skin. Therefore, the surface energy from Equation~\eqref{eq:fsurf-etf} may include some spurious bulk contribution. Nevertheless, such an extraction of $F_{\rm surf}^{\rm TETF}$ is necessary to compare with the applied CLDM.
		
\begin{figure}[H]
			\includegraphics[width=13.5 cm]{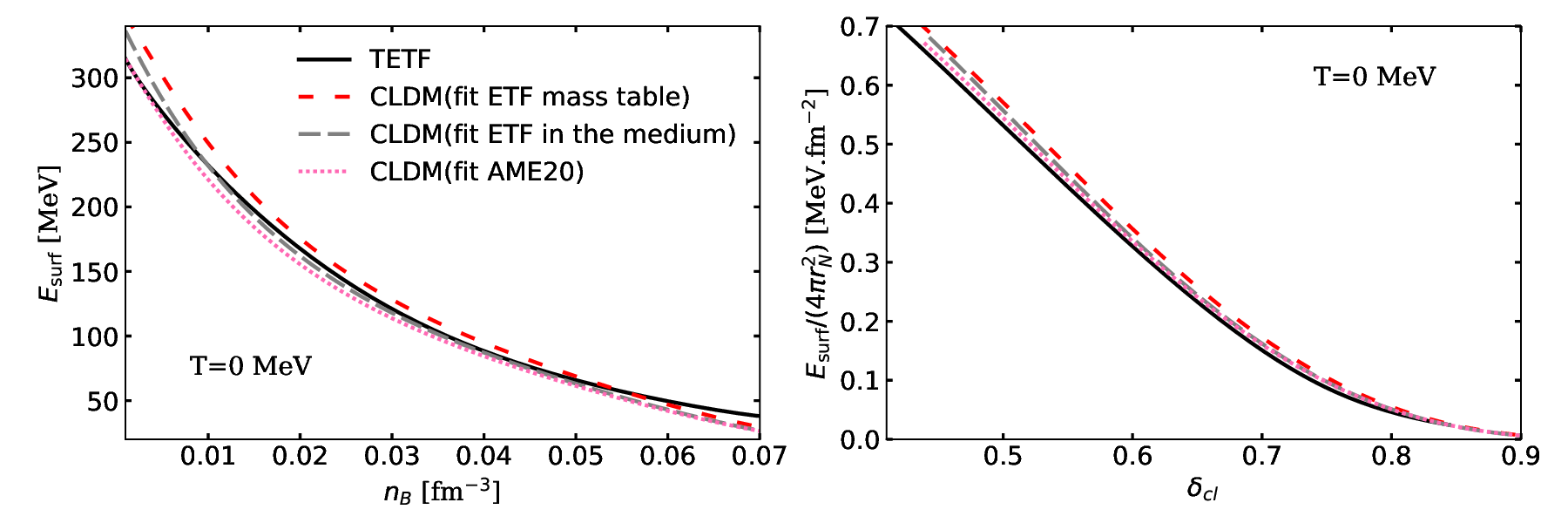}\\
			\includegraphics[width=13.5 cm]{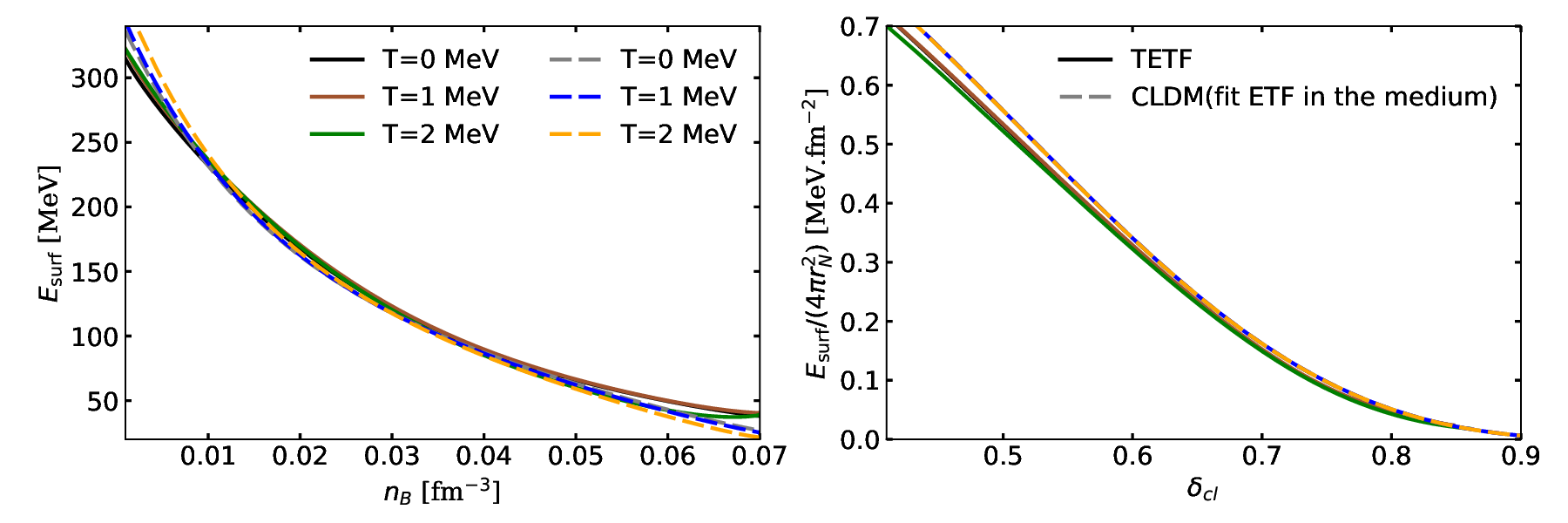}
			\caption{Total surface energy as a function of baryon density in the NS inner crust (left panels) and surface energy per unit area as a function of the cluster asymmetry $\delta_{\rm cl}$ (right panels). Results at $T=0$~MeV obtained with TETF (solid lines) and CLDM with surface parameters fitted to the ETF mass table (red dashed lines), to ETF calculations in the medium (grey dashed lines), and to the AME2020 table (pink dotted lines) are shown in the top panels.
				Results at $T =1,2$~MeV obtained with TETF (solid lines) and CLDM with surface parameters fitted to ETF calculations in the medium (dashed lines) are displayed in the bottom panels; for comparison, results at $T=0$~MeV are also shown.
			}
			
			\label{fig:Es}
		\end{figure}

		Results are shown in Figure~\ref{fig:Es}, where we display
		the surface properties of clusters in the NS inner crust as predicted by (T)ETF calculations (solid lines) and by CLDM ones obtained with the three different sets of the surface-energy parameters, as described in Section~\ref{sec:cldm} and Appendix~\ref{app:etf-fit}.
		The differences among the CLDM results arise from the different fitting procedures of the surface parameters (we remind that curvature corrections are also included; see Equations~\eqref{eq:interface}--\eqref{eq:sigma0}). 
		We show the results for the CLDM with surface parameters fitted to the ETF mass table (red dashed lines), to ETF calculations in the medium (grey dashed lines), and to the experimental masses~\cite{AME2020} (pink dotted lines).
		For all the surface-energy fits in the CLDM, the bulk empirical parameters (see Section~\ref{sec:cldm} and Equation~\eqref{eq:vMM}) are fixed to those of the BSk24 functional~\cite{Goriely13}.
		From the top left panel, we observe that at $T=0$~MeV, the CLDM with surface parameters fitted according to Appendix~\ref{app:etf-fit} indeed predicts very similar surface energies with respect to the ones extracted from Equation~\eqref{eq:fsurf-etf}.
		Meanwhile, the CLDM with surface parameters adjusted to the ETF mass tables (experimental masses) generally overestimates (underestimates) the surface energy in the inner crust.
		Comparing the calculations at finite temperature (bottom left panel), we note a reasonably good reproduction of the TETF values with the CLDM fitted to the zero-temperature ETF calculations in the medium.
		In the right panels of Figure~\ref{fig:Es}, we show the surface energy per unit area as a function of the cluster asymmetry 
		$\delta_{\rm cl} = 1 - 2~Z/A$, where for the TETF $A=4\pi n_{i}r_N^3/3$, with $n_{i}$ defined in Equation~\eqref{eq:ncl-etf}.
		We note a good agreement of the predictions of the CLDM and the (T)ETF approaches, with slightly lower values observed for the TETF, and a small impact of the temperature.
		On the other hand, one might expect a perfect agreement between the extraction of the surface energy from ETF calculations at $T=0$ and the CLDM predictions when the surface parameters are fitted to ETF calculations in the medium.
		However, as discussed in Appendix~\ref{app:etf-fit}, the CLDM fit has been performed on ETF computations \cite{pearson2018, pearson_priv} that are not the exactly the same as those used in this work; moreover, some residual Coulomb contribution could appear in the fitted surface energy. 
		These factors likely explain the small discrepancy. 
		Nevertheless, we consider the agreement observed in Figure~\ref{fig:Es} as being satisfactory.

		\begin{adjustwidth}{-\extralength}{0cm}
			\printendnotes[custom] 
			
			\reftitle{References}
			
			

			\PublishersNote{}
		\end{adjustwidth}
	\end{document}